\documentstyle[twoside,fleqn,espcrc2]{article}
\def\bfone{\relax{\rm 1\kern-.35em 1}}
\def\bfzero{\relax{\rm I\kern-.18em 0}}
\def\inbar{\vrule height1.5ex width.4pt depth0pt}
\def\IC{\relax\,\hbox{$\inbar\kern-.3em{\rm C}$}}
\def\ID{\relax{\rm I\kern-.18em D}}
\def\IF{\relax{\rm I\kern-.18em F}}
\def\IH{\relax{\rm I\kern-.18em H}}
\def\II{\relax{\rm I\kern-.17em I}}
\def\IN{\relax{\rm I\kern-.18em N}}
\def\IP{\relax{\rm I\kern-.18em P}}
\def\IQ{\relax\,\hbox{$\inbar\kern-.3em{\rm Q}$}}
\def\IR{\relax{\rm I\kern-.18em R}}
\def\IG{\relax\,\hbox{$\inbar\kern-.3em{\rm G}$}}
\font\cmss=cmss10 \font\cmsss=cmss10 at 7pt
\def\ZZ{\relax\ifmmode\mathchoice
{\hbox{\cmss Z\kern-.4em Z}}{\hbox{\cmss Z\kern-.4em Z}}
{\lower.9pt\hbox{\cmsss Z\kern-.4em Z}}
{\lower1.2pt\hbox{\cmsss Z\kern-.4em Z}}\else{\cmss Z\kern-.4em
Z}\fi}
\def\tilde{\widetilde}
\def\bar{\overline}

\def\hat{\widehat}

\def\Coe#1.#2.{{#1\over #2}}

\def\coe#1.#2.{\relax{\textstyle {#1 \over #2}}\displaystyle}

\def\to{\rightarrow}
\def\notin{\hbox{{$\in$}\kern-.51em\hbox{/}}}

\def\IE{\relax{{\rm I\kern-.18em E}}}

\def\IGam{\relax{{\rm I}\kern-.18em \Gamma}}

\def\IA{\relax{\hbox{{\rm A}\kern-.82em {\rm A}}}}

\def\o#1#2{{{#1}\over{#2}}}
\newcommand{\be}{\begin{equation}}
\newcommand{\ee}{\end{equation}}
\newcommand{\ba}{\begin{eqnarray}}
\newcommand{\ea}{\end{eqnarray}}
\newtheorem{definizione}{Definition}[section]
\newcommand{\bd}{\begin{definizione}}
\newcommand{\ed}{\end{definizione}}
\newtheorem{teorema}{Theorem}[section]
\newcommand{\bth}{\begin{teorema}}
\newcommand{\eth}{\end{teorema}}
\newtheorem{lemma}{Lemma}[section]
\newcommand{\blem}{\begin{lemma}}
\newcommand{\elem}{\end{lemma}}
\newcommand{\brr}{\begin{array}}
\newcommand{\err}{\end{array}}

\newtheorem{corollario}{Corollary}[section]
\newcommand{\bcorol}{\begin{corollario}}
\newcommand{\ecorol}{\end{corollario}}
\def\twovec#1#2{\left(\begin{array}{c}
{#1}\\ {#2}\\
\end{array}
\right)}
\newcommand{\AmS}{{\protect\the\textfont2
  A\kern-.1667em\lower.5ex\hbox{M}\kern-.125emS}}
\title{THE COMPLETE FORM of N=2 SUPERGRAVITY \\
and its PLACE in THE GENERAL FRAMEWORK \\
of D=4 N--EXTENDED SUPERGRAVITIES}
\author{Pietro
Fr\'e\address{Dipartimento di Fisica Teorica, Universit\`a di Torino,
\\ Via P. Giuria 1, I-10125 TORINO, Italy}%
\thanks{Email:
Fre@to.infn.it } }
\begin{document}
\begin{abstract}
Relying on the geometrical set up of Special K\"ahler Geometry
and Quaternionic Geometry, which I discussed at length in my Lectures
at the 1995 edition of this Spring School, I present here
the recently obtained fully general form of   N=2 supergravity
with completely arbitrary couplings. This lagrangian has
already been used in the literature to obtain various results:
notably the partial breaking of supersymmetry and various extremal
black--hole solutions. My emphasis,  however,  is only on providing the reader
with a completely explicit and ready to use component expression of the
supergravity action. All the details of the derivation are omitted
but all the definitions of the items entering the lagrangian and
the supersymmetry transformation rules are given.
\end{abstract}
\maketitle
\section{Introduction}
\label{intro}
As a consequence of the recent exciting developments on the
non--perturbative regimes of string theory,
it is by now clear that all $N$--extended supergravities in
the diverse dimensions from $D=11$ to $D=4$, constitute the low
energy effective actions for a connected
web of theories, describing  both elementary and solitonic extended objects
(the $p$--branes), that are related to each other by a
complicated pattern of
dualities, usually classified under the names of  $S$, $T$ and $U$
\cite{sumschwarz}.
These are generalizations of the electric--magnetic duality
transformations of Maxwell theory and  are just mere consequences
of the remarkable geometric structure displayed by the scalar
sector of supergravity theories. This structure has been known in the literature
for already ten to fifteen  years. In
the last year edition of the Trieste Spring School I had the
opportunity to give a series of lectures on the general form of these
electric--magnetic duality rotations and on their bearing on the
geometric structure of supergravity lagrangians \cite{mylecture}. As
I emphasized there, the key instrument to understand the full
interaction structure of supergravity theory is indeed the differential
geometry of the scalar field sector and its symplectic embedding.
Within the general framework, my main concern  was the
description of the peculiar geometric structures displayed by $N=2$
supergravity, namely the {\it Special K\"ahler Geometry} pertaining
to the vector multiplet sector and the {\it Quaternionic Geometry}
pertaining to the hypermultiplet sector. Although the focus of those
lectures was supergravity, no explicit mention was there given of the
fermion fields, the purpose being the description of the bosonic
sector geometry. In the months elapsed from last year lectures
a fully general symplectic covariant formulation of $N=2$
supergravity based on arbitrary special K\"ahler manifolds and
arbitrary quaternionic manifolds with the gauging of an arbitrary
group has become available \cite{bertolo}. This formulation includes
all previous formulations obtained both by means of the conformal tensor calculus
\cite{specspec2} and by means of the rheonomic approach \cite{skgsugra_4},
\cite{skgsugra_1} but extends them to the most general situation.
Notably within this formulation one can accommodate cases that were
out of reach of previous formulations (for instance those where the special
K\"ahler geometry admits no prepotential $F(X)$) and which are
physically particularly relevant. Applications of this
new formulation  have already appeared in the literature. In particular
in \cite{mariotrig}, by extending results obtained in \cite{maxpor}, it
has been shown that $N=2$ supersymmetry can spontaneously break to $N=1$
with a surviving unbroken compact gauge group.
\par
In the present seminar it is my purpose to present the complete
most general form of the N=2 supergravity lagrangian and of the
supersymmetry transformation rules against which it is invariant.
\section{How N=2 supergravity fits in the general scheme}
Restricting my attention to the $D=4$ theories, I can recall from
\cite{mylecture} that in their bosonic part all $N$--extended
supergravity lagrangians admit the following general form:
\begin{eqnarray}
&{\cal L}^{SUGRA}_{Bose} = & \nonumber\\ & \sqrt{-g}
\, \Bigl ( \, R \, + \, g_{IJ}(\phi) \, \nabla_\mu \phi^I \,
\nabla^\mu \phi^J & \nonumber\\ &+ \, {\rm i} \, \left( \bar {\cal
N}_{\Lambda \Sigma} {\cal F}^{- \Lambda}_{\mu \nu} {\cal F}^{-
\Sigma \vert {\mu \nu}} \, - \, {\cal N}_{\Lambda \Sigma} {\cal
F}^{+ \Lambda}_{\mu \nu} {\cal F}^{+ \Sigma \vert {\mu \nu}} \right
)& \nonumber \\ & - {\cal V} (\phi) \Bigl ) & \label{genform}
\end{eqnarray}
where $\phi^I$ denote the $n_S$ scalar fields of the
theory and
\begin{equation} {\cal F}^{\pm \Lambda}_{\mu \nu} =
{1\over 2} \, \Bigl ( F^{\Lambda}_{\mu \nu} \pm {1\over 2}
\epsilon_{\mu\nu\rho\sigma}F^{\Lambda \vert \mu \nu} \Bigr )
\label{dualcombo}
\end{equation}
denote the (anti)self--dual
combination of the $n_V$ field strengths:
\begin{equation} {\cal
F}^{ \Lambda}_{\mu \nu} \equiv {1 \over 2} \Bigl ( \partial_{\mu}
A^{\Lambda}_\nu -\partial_\nu A^{\Lambda}_\mu + g
f^{\Lambda}_{\Sigma\Delta} \, A^{\Sigma}_\mu \, A^{\Delta}_\nu \Bigr
)
\end{equation}
Indeed, whether pure or matter coupled, each
supergravity theory is primarily characterized by these two
numbers, {\it i.e.}
the total number of scalars and the total number of vector fields.
One has:
\begin{eqnarray}
n_S &=& \# \mbox{ scal. in scal. mult.} \, \nonumber\\
&& + \, \# \mbox{ scal. in vect. mult.} \, \nonumber\\
&& + \, \# \mbox{ scal. in grav. mult.} \, \nonumber\\
n_V &=& \# \mbox{ scal. in vect. mult.} \, \nonumber\\
&& + \, \# \mbox{ scal.
in grav. mult.}
\end{eqnarray}
and the available choices are summarized
in table \ref{topotable}.
To continue the illustration of eq.~\ref{genform}, by $\nabla_\mu
\phi^I$ I denote the derivatives of the scalar fields, covariant
with respect to the action of the gauge group $G_{gauge}$:
\begin{equation}
\nabla_\mu \phi^I = \partial \phi^I + g \,
A^\Lambda_\mu \, k^I_\Lambda (\phi)
\label{kilvec}
\end{equation}
The Killing vector fields, appearing in eq.\ref{kilvec}
\begin{equation}
{ \vec {\bf k}}_\Lambda \, \equiv \, k^I_\Lambda
(\phi)\, {\o{\partial}{\partial \phi^I}}
\end{equation}
satisfy the Lie algebra ${\bf G}_{gauge}$ of the gauge group:
\begin{equation}
\Bigr [ { \vec {\bf k}}_\Lambda \, , \, { \vec {\bf k}}_\Sigma \,
\Bigr ] \, = \, f^\Delta_{\Lambda\Sigma} \, { \vec {\bf k}}_\Delta
\end{equation}
whose dimension is less or equal to the number of
vector fields $n_V$. This algebra is a subalgebra of the isometry
algebra of the scalar metric:
\begin{equation} {\bf G}_{gauge} \,
\subset \, {\bf G}_{iso}
\end{equation}
By definition ${\bf G}_{iso}$ is generated by all those vector fields
${\vec {\bf t}} \, \in \, T{\cal M}_{scalar}$ such that:
\begin{equation}
\ell_{{ \vec {\bf t}} } \, g_{IJ} = 0
\label{invariant}
\end{equation}
the operator $\ell$ denoting the Lie derivative.
Hence eq.\ref{invariant} is in particular true when
${\vec {\bf t}}$ = ${\vec {\bf k}}_\Lambda$.
As extensively discussed in \cite{mylecture}, the complex symmetric
matrix ${\cal N}_{\Lambda\Sigma}(\phi)$, whose real and imaginary
part  play, respectively, the role of field dependent $theta$--angle
and field dependent coupling constant,  is determined by the
symplectic embedding of the isometry group $G_{iso}$:
\begin{equation}
\iota_\delta \, : \, G_{iso} \,  \longrightarrow \, Sp(2\,n_V, \IR)
\label{iota}
\end{equation}
so that for each element $\xi \, \in \, G_{iso}$ we have:
\begin{eqnarray}
 &\xi    :     {\cal M}_{scalar} \, \longrightarrow \,{\cal M}_{scalar} & \nonumber\\
  &\forall {\vec X}, {\vec Y}  \in   T{\cal M}_{scalar} :& \nonumber\\
 & g \Bigl (\xi^\star {\vec X} ,\xi^\star {\vec Y } \Bigl ) =
 g \Bigl ( {\vec X} ,  {\vec Y } \Bigl ) &  \nonumber\\
 & {\cal N}\left (\xi \left ( \phi \right ) \right )    =
 \left ( C_\xi +D_\xi {\cal N} \right) \left ( A_\xi + B_\xi  {\cal N} \right)^{-1}&
 \nonumber \\
 \end{eqnarray}
 where
 \begin{equation}
 \iota_\delta \left ( \xi \right ) \, = \, \left ( \matrix { A_\xi & B_\xi \cr
 C_\xi & D_\xi \cr } \right ) \, \in \, Sp(2 n_V, \IR ) \nonumber\\
 \end{equation}
 is the symplectic image of $\xi$.
 When the scalar manifold is a homogeneous symmetric coset manifold, as it happens
 in all $N\ge 3$ supergravities (see Table \ref{topotable}), the matrix ${\cal N}_{\Lambda\Sigma}$
 is determined by a universal formula that was derived long ago by
 Gaillard and Zumino \cite{gaizum}. Taking into account the
 isomorphism:
\begin{eqnarray}
& Sp(2 n_V, \IR)  \sim  Usp(n_V , n_V)   \equiv& \nonumber\\
& Sp(2 n_V, \IC)   \cap   U(  n_V ,   n_V) &
\label{usplet}
\end{eqnarray}
that is explicitly realized by the map:
\begin{eqnarray}
&\mu \, : \, Sp(2 n_V, \IR )\, \ni \, \left ( \matrix { A  & B  \cr
 C  & D \cr } \right ) \,  \longrightarrow & \nonumber\\
  & \left ( \matrix{ T & V^\star \cr V & T^\star \cr } \right ) \, \in \, Usp(n_V,n_V)&
\label{blocusplet}
\end{eqnarray}
where:
\begin{eqnarray}
T &=& {\o{1}{2}}\, \left ( A - {\rm i} B \right ) +
{\o{1}{2}}\, \left ( C + {\rm i} D \right ) \nonumber\\
V &=& {\o{1}{2}}\, \left ( A - {\rm i} B \right ) -
{\o{1}{2}}\, \left ( C + {\rm i} D \right )
\label{mappetta}
\end{eqnarray}
the symplectic embedding is such that, for $M_{scalar}$ = $G_{iso}/H$ we
have:
\begin{eqnarray}
\mu \, \cdot \, \iota_\delta \left ( G_{iso} \right) & \subset & Usp(n_V,n_V) \nonumber\\
 \mu \, \cdot \, \iota_\delta \left ( H \right) & \subset & U(n_V) \, \subset \, Usp(n_V,n_V)
\end{eqnarray}
Then, considering an arbitrary parametrization of the coset  $G_{iso}/H$ {\it i.e.}:
\begin{equation}
 G_{iso}/H \, \ni \, \phi \, \longrightarrow \, L(\phi) \, \in \, G
 \end{equation}
by setting:
 \begin{eqnarray}
 & Usp(n_V,n_V)\, \ni \, {\cal O}(\phi) \, \equiv \, \mu \, \cdot \,
 \iota_\delta \left ( L(\phi) \right ) = & \nonumber\\
 & =    \left ( \matrix{ U_0(\phi) & U^\star_1(\phi) \cr U_1(\phi)
& U^\star_0(\phi) \cr } \right ) &
\end{eqnarray}
we immediately obtain an immersion of $G_{iso}/H$ into
$Usp(n_V,n_V)/U(n_V)$ and the matrix ${\cal N}$ is obtained by defining,
according to \cite{gaizum}:
\begin{equation}
{\cal N} \, \equiv \, {\rm i} \left [ U_0^\dagger + U_1^\dagger \right
]^{-1} \, \left [ U_0^\dagger - U_1^\dagger \right ]
\label{masterformula}
\end{equation}
The peculiarity of $N=1$ and $N=2$ supergravity with respect to the
higher theories is that the scalar manifold is not necessarily  a
homogeneous symmetric coset manifold. Indeed the only request
imposed by $N=2$ supersymmetry on the scalar manifold $M_{scalar}$
is that it should be the direct product:
\begin{equation}
M_{scalar} \, = \, {\cal SM}_n \, \otimes \, {\cal HM}_m
\end{equation}
of a special K\"ahler manifold ${\cal SK}_n$ of complex dimension
$\mbox{dim}_{\bf C} \, {\cal SM}_n =n$ equal to the number of
vector multiplets with a quaternionic manifold of quaternionic
dimension $\mbox{dim}_{\bf Q} \,{\cal HM}_m \, = \,m$  equal to the
number of hypermultiplets. Correspondingly the scalar metric is of
the form:
\begin{eqnarray}
g_{IJ}(\phi)\, d\phi^I \otimes d\phi^J \,& \equiv &\, g_{ij^\star}\, dz^i \otimes
d{\bar z}^{j^\star} \nonumber\\
&& + \, h_{uv} \, dq^u \, \otimes \,dq^v
\end{eqnarray}
where $g_{ij^\star}$ is the special K\"ahler metric on ${\cal SM}_n$
and $h_{uv}$ is the quaternionic metric on ${\cal HM}_m$. Following
the discussion of \cite{mylecture} there exists a symplectic holomorphic vector
bundle with structural group $Sp(2n+2,\IR)$ and a section
\begin{equation}
\Omega = \left ( \matrix{X^\Lambda \cr F_\Sigma \cr} \right )
\end{equation}
such that the K\"ahler potential can be written as:
\begin{eqnarray}
{\cal K}\,  &=& \,  -\mbox{log}\left ({\rm i}\langle \Omega \,
 \vert \, \bar \Omega
\rangle \right )\,\nonumber \\
&&=\, -\mbox{log}\left [ {\rm i} \left ({\bar X}^\Lambda F_\Lambda -
{\bar F}_\Sigma X^\Sigma \right ) \right ]
\label{specpot}
\end{eqnarray}
Defining furthermore
\begin{equation}
V \, = \, \twovec{L^{\Lambda}}{M_\Sigma} \, \equiv \, e^{{\cal K}/2}\Omega
\,= \, e^{{\cal K}/2} \twovec{X^{\Lambda}}{F_\Sigma}
\label{covholsec}
\end{equation}
one has
\begin{equation}
\nabla_{i^\star} V \, = \, \left ( \partial_{i^\star} - {\o{1}{2}}
\partial_{i^\star}{\cal K} \right ) \, V \, = \, 0
\label{nonsabeo}
\end{equation}
and setting:
\begin{equation}
U_i   =   \nabla_i V  =   \left ( \partial_{i} + {\o{1}{2}}
\partial_{i}{\cal K} \right ) \, V   \equiv
\twovec{f^{\Lambda}_{i} }{h_{\Sigma\vert i}}
\label{uvector}
\end{equation}
it follows that:
\begin{equation}
\nabla_i U_j  = {\rm i} C_{ijk} \, g^{k\ell^\star} \, {\bar U}_{\ell^\star}
\label{ctensor}
\end{equation}
where $\nabla_i$ denotes the covariant derivative containing both
the Levi--Civita connection on the bundle ${\cal TM}$ and the
canonical connection $\theta$ on the line bundle ${\cal L}$
whose first Chern class equal the K\"ahler class.
In eq.~\ref{ctensor} the symbol $C_{ijk}$ denotes a covariantly
holomorphic (
$\nabla_{\ell^\star}C_{ijk}=0$) section of the bundle
${\cal TM}^3\otimes{\cal L}^2$ that is totally symmetric in its indices.
It enters in the construction of the lagrangian and of the
transformation rules together with the upper part $L^\Lambda$ of the
non--holomorphic symplectic section $V$ and the upper part of its
derivative $f^\Lambda_i$. The matrix ${\cal N}$ is defined by the
identity:
\begin{equation}
{\bar M}_\Lambda = {\bar {\cal N}}_{\Lambda\Sigma}{\bar L}^\Sigma \quad ;
\quad
h_{\Sigma\vert i} = {\bar {\cal N}}_{\Lambda\Sigma} f^\Sigma_i
\label{etamedia}
\end{equation}
In the $N=2$ case, the Killing vector ${\vec {\bf k}}_\Lambda$ is
composed of two parts.
\begin{equation}
  {\vec {\bf k}}_\Lambda = \left [k_\Lambda^i \, {\o{\partial}{\partial z^i}}
  + k_\Lambda^{i^\star} \, {\o{\partial}{\partial z^{i^\star}}}
  \right ]
  +k_\Lambda^u {\o{\partial}{\partial q^{u}}}
\end{equation}
corresponding to the infinitesimal action of the gauge group on the
special K\"ahler manifold and on the quaternionic manifold,
respectively. Naming $K^x$ the triplet of HyperK\"ahler 2--forms
that, by definition of quaternionic manifold, do necessarily exist
on ${\cal HM}$, we introduce the notion of triholomorphic momentum
map  associating to each killing vector
$k_\Lambda^u {\o{\partial}{\partial q^{u}}}$ a triplet of
0--form prepotentials ${\cal P}^z_\Lambda$ defined through the
following equation:
\begin{equation}
{\bf i}_\Lambda  K^x =
- \nabla {\cal P}^x_\Lambda \equiv -(d {\cal
P}^x_\Lambda + \epsilon^{xyz} \omega^y {\cal P}^z_\Lambda)
\label{2.76}
\end{equation}
The components of the HyperK\"ahler 2--forms satisfy the
quaternionic algebra:
\begin{equation}
h^{st} K^x_{us} K^y_{tw} = -   \delta^{xy} h_{uw} +
  \epsilon^{xyz} K^z_{uw}
\label{universala}
\end{equation}
In addition one  needs also the quaternionic vielbein ${\cal
U}^{A\alpha}$ that is a 1--form with a pair of tangent indices,
the first ($A$ ) taking two values and transforming in the doublet
representation of $SU(2)$, the second ($\alpha$) taking $2m$ values
and transforming in the fundamental representation of $Sp(2m)$.
The relation between the quaternionic vielbein, the quaternionic
metric and the triplet of HyperK\"ahler 2--forms is given by:
\begin{eqnarray}
({\cal U}^{A\alpha}_u {\cal U}^{B\beta}_v + {\cal U}^{A\alpha}_v {\cal
 U}^{B\beta}_u)\IC_{\alpha\beta}&=& h_{uv} \epsilon^{AB}\nonumber\\
({\cal U}^{A\alpha}_u {\cal U}^{B\beta}_v + {\cal U}^{A\alpha}_v {\cal
U}^{B\beta}_u) \epsilon_{AB} &=& h_{uv} {1\over m} \IC^{\alpha
\beta} \nonumber\\
\, {\rm i}\,   \IC_{\alpha\beta}
(\sigma _x)_A^{\phantom {A}C}\epsilon _{CB} {\cal U}^{\alpha A} \wedge {\cal U}^{\beta B}
&=& K^x
\label{piuforte}
\end{eqnarray}
where
\begin{equation}
{\cal U}_{A\alpha} \equiv ({\cal U}^{A\alpha})^*  =  \epsilon_{AB}
\IC_{\alpha\beta} {\cal U}^{B\beta}
\end{equation}
and where $\epsilon_{AB}$ denotes the Levi--Civita tensor in
2--dimensions and $\IC_{\alpha\beta}$ is the symplectic invariant
constant antisymmetric matrix.
\par
To complete the list of geometrical
data entering the supergravity lagrangian one still needs the
$SU(2) \otimes Sp(2m)$ Lie algebra valued spin connection of
${\cal HM}$ that is implicitly defined by the vanishing torsion
equation:
\begin{eqnarray}
\nabla {\cal U}^{A\alpha}& \equiv & d{\cal U}^{A\alpha}
+{i\over 2} \omega^x (\epsilon \sigma_x\epsilon^{-1})^A_{\phantom{A}B}
\wedge{\cal U}^{B\alpha} \nonumber\\
&+& \Delta^{\alpha\beta} \wedge {\cal U}^{A\gamma} \IC_{\beta\gamma}
=0
\label{quattorsion}
\end{eqnarray}
\section{The fermion fields}
In order to present the full supergravity lagrangian we need to
introduce the fermion fields. They are
\begin{itemize}
\item {the gravitino:
\begin{equation}
\psi_A \, \equiv \,  \psi_{A\mu} \, dx^\mu
\end{equation}
which is an $SU(2)$ doublet of chiral ( $\gamma_5 \, \psi_A =
\psi_A$)
spinor valued 1--forms (the complex conjugate doublet corresponding
to the opposite chiral projection : $\gamma_5 \, \psi_A =
\psi_A$)}
\item{ the gauginos:
\begin{equation}
\gamma_5 \,  \lambda^{iA} = \lambda^{iA}  \quad ; \quad \gamma_5 \,  \lambda^{i^\star}_A = -
\lambda^{i^\star}_A
\end{equation}
which, besides being 4--dimensional spinors, transform as world tensors
on the special K\"ahler manifold and as sections of the Hodge line--bundle}
\item{ the hyperinos
\begin{equation}
 \gamma_5 \,  \zeta^\alpha = - \zeta^\alpha  \quad ; \quad \gamma_5 \,  \zeta_\alpha =
 \zeta_\alpha
\end{equation}
 which, besides being 4--dimensional spinors, transform in the fundamental representation
 of the $Sp(2m)$ holonomy group of ${\cal HM}$}
\end{itemize}
Then the covariant derivatives of the fermion fields are:
\begin{eqnarray}
\rho_A &  \equiv & d\psi_A-{1\over 4} \gamma_{ab} \,
\omega^{ab}\wedge\psi_A \nonumber\\
&&+ {{\rm i} \over 2} {\hat {\cal Q}}\wedge \psi_A +
{\hat \omega}_A^{~B}\wedge \psi_B
 \nonumber\\
\rho^A & \equiv & d\psi^A-{1\over 4} \gamma_{ab} \, \omega^{ab}\wedge\psi^A \nonumber\\
&&-{{\rm i} \over 2} {\hat {\cal Q}}\wedge\psi^A
+{\hat \omega}^{A}_{\phantom{A}B} \wedge \psi^B
 \nonumber\\
 \nabla\lambda^{iA} &\equiv & d\lambda^{iA}-{1\over 4} \gamma_{ab} \,
\omega^{ab} \lambda^{iA} \nonumber\\
&&-{{\rm i} \over 2} {\hat {Q}}\lambda^{iA}+
{\hat \Gamma}^i_{\phantom{i}j}\lambda^{jA}+{\hat \omega}^{A}_{~B} \wedge
\lambda^{iB}
\nonumber\\
\nabla\lambda^{{i}^\star}_A &\equiv &d\lambda^{\bar
{\imath}}_A-{1\over 4} \gamma_{ab} \,
\omega^{ab}\lambda^{{i}^\star}_A \nonumber\\
&&+{{\rm i} \over 2}
{\hat{\cal Q}}\lambda^{{i}^\star}_A+
{\hat \Gamma}^{{i}^\star}_{\phantom{\bar
{\imath}}{{j}^\star}}\lambda^{{j}^\star}_A
+{\hat \omega}_{A}^{~B} \wedge
\lambda^{{{i}^\star}}_B
\nonumber\\
\nabla \zeta _{\alpha} & \equiv & d \zeta _{\alpha} \,-\,{1 \over 4}
\omega ^{ab}\,\gamma _{ab}\,\zeta _{\alpha}\nonumber\\
&&-{{\rm i} \over 2} {\hat {\cal Q}}\,\zeta _{\alpha}\,+
{\hat {\Delta}}_{\alpha}^{\phantom{\alpha}{\beta}}\zeta _{\beta}
\nonumber\\
\nabla \zeta^{\alpha} & \equiv & d \zeta^{\alpha} \,-\,{1 \over 4}
\omega ^{ab}\,\gamma _{ab}\,\zeta^{\alpha}\nonumber\\
&&+{{\rm i} \over 2} {\hat {\cal Q}}\,\zeta^{\alpha}\,+
{\hat {\Delta}}^{\alpha}_{\phantom{\alpha}{\beta}}\zeta^{\beta}
\end{eqnarray}
In the above definitions the connections acting on the fermion fields
are the gauged ones, namely:
\begin{equation}
\begin{array}{ ccc}
\Gamma^{i}_{\phantom{i}j}& \to &{\hat \Gamma}^{i}_{\phantom{i}j} =
 \Gamma^{i}_{\phantom{i}j} +
 g\, A^\Lambda\, \partial_j k^i_\Lambda \cr
{\cal Q} &\to &{\hat {\cal Q}}= {\cal Q} - i g\, A^\Lambda\,  k^i_\Lambda \,
\partial_i {\cal K}\cr
\omega^x &\to &{\hat \omega}^x = \omega^x + g\, A^\Lambda\, {\cal
P}^x_\Lambda \cr
\Delta^{\alpha\beta} &\to &{\hat  \Delta}^{\alpha\beta}=
\Delta^{\alpha\beta}  \cr
\null & \null & + g\, A^\Lambda\,
 \partial_u k_\Lambda^v \, {\cal U}^{u \vert  \alpha A}
 \, {\cal U}^\beta_{v \vert A} \cr
 \end{array}
\label{compogauging}
\end{equation}
%%%%%%%%%%%%%%
\section{The Lagrangian and the supersymmetry transformation rules}
In terms of the geometrical data so far introduced and of the
covariant derivatives defined above the complete N=2 supergravity
Lagrangian can now be presented. It is displayed in table 2.
To make the formula completely explicit it suffices to recall
that by $(...)^-$ I have denoted the self dual part of
any fermion bilinear combination involving the $\gamma_{\mu\nu}$ matrix.
Furthermore the explicit form of the fermion mass--matrices  appearing in the lagrangian
is given below:
\begin{center}
\begin{tabular}{c}
\null\\
\hline
\null \\
{\it  N=2 Supergravity mass matrices}\\
\null \\
\hline
\end{tabular}
\end{center}
\vskip 0.2cm
\begin{eqnarray}
S_{AB}&=&{{\rm i} \over 2} (\sigma_x)_A^{\phantom{A}C} \epsilon_{BC}
{\cal P}^x_{\Lambda}L^\Lambda \nonumber\\
W^{iAB}&=&\epsilon^{AB}\,k_{\Lambda}^i \bar L^\Lambda\,+ \nonumber\\
& &
{\rm i}(\sigma_x)_{C}^{\phantom{C}B} \epsilon^{CA} {\cal P}^x_{\Lambda}
g^{i{{j}^\star}} {\bar f}_{{j}^\star}^{\Lambda}\nonumber\\
N^A_{\alpha}&=& 2 \,{\cal U}_{\alpha \vert u}^A \,k^u_{\Lambda}\,\bar L^{\Lambda}\nonumber\\
{\cal M}^{\alpha \vert \beta}  &=&-
\, g \, {\cal U}^{\alpha A}_u \, {\cal U}^{\beta B}_v \, \varepsilon_{AB}
\, \nabla^{[u}   k^{v]}_{\Lambda}  \, L^{\Lambda} \nonumber\\
{\cal M}^{\alpha\vert }_{\phantom{\alpha}\vert iB} &=&-
4 \, g \, {\cal U}^{\alpha}_{B \vert u} \, k^u_{\Lambda} \, f^{\Lambda}_i \nonumber\\
{\cal M}_{iA\vert \ell B} &=& \,{\o{1}{3}} \, g \,
\Bigl ( \varepsilon_{AB}\,  g_{ij^\star}   k^{j^\star}_ \Lambda   f_\ell^\Lambda \nonumber\\
& &+
{\rm i}\bigl (\sigma_x)_A^{\phantom{A}C} \epsilon_{BC} \, {\cal P}^x_ \Lambda
\, \nabla_\ell f^\Lambda _i \Bigr )
\label{pesamatrice}
\end{eqnarray}
The coupling constant in front of the mass--matrices and of the
potential is just a symbolic notation to remind the reader
that these terms are entirely
due to the gauging and vanish in the ungauged theory. In general
there is not a single coupling constant rather there are
as many independent coupling constants as the number of factors in
the gauge group. This fact has been exploited in
\cite{mariotrig,maxpor} to obtain the partial supersymmetry breaking
$N=2 \, \longrightarrow \, N=1$ by gauging a non compact abelian
group with appropriate ratio of the coupling constants associated
with each generator.
%%%%%%%%%%%%%%%%%%%%%%%%%%%%%%%%%%%%%
%%leggi di trasformazione dei campi%%
%%%%%%%%%%%%%%%%%%%%%%%%%%%%%%%%%%%%%
\par
\section{The supersymmetry transformation rules}
Of prominent interest in many applications is, besides that of the action,
the form of the supersymmetry transformation rules.
For instance this information is  essential in order to obtain the differential
equations describing BPS saturated states \cite{kalvanp1},\cite{ferkal2},\cite{strom3},
\cite{ferkal4},\cite{cardoso},
namely those field
configurations that preserve $1/2$ of the original supersymmetries.
Another respect where the supersymmetry transformation rules
are of vital importance is in the topological twist to topological
field theories \cite{topftwist_1}, \cite{topftwist_2}, \cite{topf4d_8},
\cite{noi}.
\par
The explicit form of the SUSY rules is listed below
\begin{center}
\begin{tabular}{c}
\null\\
\hline
\null \\
{\it Supergravity transformation rules}\\
{\it of the Fermi  fields}\\
\null \\
\hline
\end{tabular}
\end{center}
\vskip 0.2cm
\begin{eqnarray}
&\delta\,\Psi _{A \vert \mu}  =& \nonumber \\
&{\cal D}_{\mu}\,\epsilon _A\,
 -{1 \over 4} \left(\partial _i\,{\cal K} \bar {\lambda}^{iB}\epsilon _B\,-\,
 \partial _{{i}^\star}\,{\cal K} \bar {\lambda}^{{i}^\star}_B \epsilon^B \right)
 \Psi _{A \vert \mu} & \nonumber\\
& -{\omega}_{A \vert v}^{\phantom{A}B}\,
{\cal U}_{C \alpha}^v\,
\left(\epsilon^{CD}\,\IC^{\alpha  \beta}\,\bar
{\zeta}_{\beta}\,\epsilon _D\,+\,
\,\bar {\zeta}^{\alpha}\,\epsilon^C \right)\Psi _{B \vert \mu} & \nonumber\\
&  +\,\left ( A_{A}^{\phantom{A} \vert{\nu}B}
\eta_{\mu \nu}+A_{A}^{\prime \phantom{A} \vert {\nu} B}\gamma_{\mu \nu} \right ) \epsilon _B \,
& \nonumber\\
&  + \left [ {\rm i} \, g \,S_{AB}\eta _{\mu \nu}+
\epsilon_{AB}( T^-_{\mu \nu}\, + \,U^+_{\mu \nu} )
\right ] \gamma^{\nu}\epsilon^B & \nonumber \\
\label{trasfgrav}
\end{eqnarray}
\begin{eqnarray}
& \delta\,\lambda^{iA} =& \nonumber\\
&{1 \over 4} \left(\partial _j\,{\cal K} \bar {\lambda}^{jB}\epsilon _B\,-\,
\partial _{{j}^\star}\,{\cal K}
\bar {\lambda}^{{j}^\star}_B \epsilon^B \right)\lambda^{iA}& \nonumber\\
& -{\omega}^{A}_{\phantom{A}B \vert v}\,
{\cal U}_{C \alpha}^v\,
\left(\epsilon^{CD}\,\IC^{\alpha  \beta}\,\bar
{\zeta}_{\beta}\,\epsilon _D\,+\,
\,\bar {\zeta}^{\alpha}\,\epsilon^C \right)\lambda^{iB} & \nonumber\\
& -\,\Gamma^i_{\phantom{i}{jk}}
\bar {\lambda}^{kB}\epsilon _B\,\lambda^{jA}
+ &\nonumber\\
& {\rm i}\, \left (\nabla _ {\mu}\, z^i -\bar {\lambda}^{iA}\psi
_{A \vert \mu}\right)
\gamma^{\mu} \epsilon^A & \nonumber\\
& +G^{-i}_{\mu \nu} \gamma^{\mu \nu} \epsilon _B \epsilon^{AB}\,+\,
D^{iAB}\epsilon _B &
\nonumber\\
 \label{gaugintrasfm}
\end{eqnarray}
\begin{eqnarray}
& \delta\,\zeta _{\alpha} =& \nonumber \\
& -\Delta _{\alpha \vert v}^{\phantom{\alpha}\beta}\,
{\cal U}_{\gamma A}^v\,
\left(\epsilon^{AB}\,\IC^{\gamma \delta}\,\bar
{\zeta}_{\delta}\,\epsilon _B\,+\,
\,\bar {\zeta}^{\gamma}\,\epsilon^A \right)\zeta _{\beta} & \nonumber\\
& + {1 \over 4} \Bigl ( \partial _i\,{\cal K} \bar {\lambda}^{iB}
\epsilon _B\,-\,
 \partial _{{i}^\star}\,{\cal K} \bar {\lambda}^{{i}^\star}_B
 \epsilon^B \Bigr )\zeta _{\alpha} & \nonumber\\
 & +\, {\rm i}\,\Bigl (
{\cal U}^{B \beta}_{u}\, \nabla _{\mu}\,q^u\, & \nonumber\\
& -\epsilon^{BC}\,\IC^{\beta\gamma}\,\bar
{\zeta}_{\gamma}\,\psi _C\,-\,
\,\bar {\zeta}^{\beta}\,\psi^B
\Bigr ) \,\gamma^{\mu} \epsilon^A
\epsilon _{AB}\,\IC_{\alpha  \beta}
\,& \nonumber\\
& + \,g\,N_{\alpha}^A\,\epsilon _A & \nonumber\\
\label{iperintrasf}
\end{eqnarray}
\begin{center}
\begin{tabular}{c}
\null\\
\hline
\null \\
{\it Supergravity transformation rules} \\
{\it of the Bose  fields}\\
\null \\
\hline
\end{tabular}
\end{center}
\vskip 0.2cm
\begin{eqnarray}
\delta\,V^a_{\mu}&=&-{\rm i}\,\bar {\Psi}_{A \vert
\mu}\,\gamma^a\,\epsilon^A -{\rm i}\,\bar {\Psi}^A _
\mu\,\gamma^a\,\epsilon_A
\nonumber \\
\delta\,A^\Lambda _{\mu} &=& 2 \bar L^\Lambda \bar \psi _{A \vert \mu} \epsilon _B
\epsilon^{AB}
  +\,2L^\Lambda\bar\psi^A_{\mu}\epsilon^B \epsilon
_{AB}\nonumber\\
&&+\Bigl ( {\rm i} \,f^{\Lambda}_i \,\bar {\lambda}^{iA}
\gamma _{\mu} \epsilon^B \,\epsilon _{AB} \nonumber\\
&&+{\rm i} \,{\bar f}^{\Lambda}_{{i}^\star} \,\bar\lambda^{{i}^\star}_A
\gamma _{\mu} \epsilon_B \,\epsilon^{AB}\Bigr )\nonumber\\
\delta\,z^i &=& \bar{\lambda}^{iA}\epsilon _A \nonumber \\
\delta\,z^{{i}^\star} &=& \bar{\lambda}^{{i}^\star}_A \epsilon^A
\nonumber  \\
  \delta\,q^u &=& {\cal U}^u_{\alpha A} \left(\bar {\zeta}^{\alpha}
  \epsilon^A + \IC^{\alpha  \beta}\epsilon^{AB}\bar {\zeta}_{\beta}
  \epsilon _B \right) \nonumber\\
 \end{eqnarray}
 In the above rules there appear certain field combinations which
 have been given special names. In an off--shell formulation of
 supergravity these would be auxiliary fields. Their explicit
 on--shell expression in terms of the fundamental physical fields is
 given below:
\begin{center}
\begin{tabular}{c}
\null\\
\hline
\null \\
{\it Supergravity values}\\
{\it of the auxiliary  fields}\\
\null \\
\hline
\end{tabular}
\end{center}
\vskip 0.2cm
\begin{eqnarray}
A_{A}^{\phantom{A} \vert \mu B}
&=&-{{\rm i} \over 4}\, g_{{{k}^\star}\ell}\,
\Bigl( \bar {\lambda}^{{k}^\star}_A \gamma^{\mu} \lambda^{\ell B}\,- \nonumber\\
&& \delta^B_A\,
\bar\lambda^{{k}^\star}_C \gamma^{\mu} \lambda^{\ell C}\Bigr )\nonumber  \\
A_{A}^{\prime \phantom{A} \vert \mu B}
&=&{{\rm i} \over 4}\, g_{{{k}^\star}\ell}\,\Bigl (
\bar{\lambda}^{{k}^\star}_A \gamma^{\mu}
\lambda^{\ell B} - \nonumber \\
&& {1\over 2}\, \delta^B_A\, \bar\lambda^{{k}^\star}_C
\gamma^{\mu} \lambda^{C\ell}\Bigr ) \,-\,  {{\rm i} \over 4}\,  \delta _A^B \,\bar \zeta _{\alpha}
\gamma^{\mu} \zeta^{\alpha}\nonumber \\
\end{eqnarray}
\begin{eqnarray}
 T^-_{\mu\nu}  &=&
\left ({\cal N}-{\bar {\cal N}}\right)_{\Lambda\Sigma} L^{\Sigma}
\Bigl ({\tilde{F}}_{\mu\nu}^{\Lambda -} \nonumber\\
&& +{1\over 8}
\nabla_i \,f^{\Lambda} _j \,
\bar \lambda^{i A} \gamma_{\mu\nu} \, \lambda^{jB} \,\epsilon_{AB}
\nonumber\\
&&-{1\over 4} \, \IC^{\alpha  \beta}\,
{\bar\zeta}_{\alpha}\gamma _{\mu\nu} \,
\zeta _{\beta}\, L^{\Lambda}
\Bigr ) \nonumber  \\
T^+_{\mu\nu}  &=&
\left ({\cal N}-{\bar {\cal N}}\right)_{\Lambda\Sigma} {\bar L}^{\Sigma}
\Bigl( {\tilde{F}}_{\mu\nu}^{\Lambda +} \nonumber \\
&& +{1\over 8} \nabla_{{i}^\star} \,\bar
f^\Lambda_{{j}^\star}\,
\bar \lambda^{{i}^\star}_A \gamma _{\mu\nu} \, \lambda^{{j}^\star}_B \epsilon^{AB}
\nonumber\\
&&-{1 \over 4}\,  \IC_{\alpha  \beta}\,{\bar\zeta}^{\alpha}\gamma _{\mu\nu} \,
\zeta ^{\beta}\, {\bar L}^{\Lambda}
\Bigr )
\nonumber \\
U^-_{\mu\nu} &=& -{{\rm i} \over 4}  \, \IC^{\alpha  \beta}\,{\bar\zeta}_{\alpha}\gamma _{\mu\nu} \,
\zeta _{\beta}\nonumber\\
\\
U^+_{\mu\nu} &=& -{{\rm i} \over 4} \,  \IC_{\alpha  \beta}\,{\bar\zeta}^{\alpha}\gamma _{\mu\nu} \,
\zeta^{\beta}\nonumber\\
 G^{i-}_{\mu\nu} &=& {{\rm i} \over 2}\,g^{i{{j}^\star}} \bar f^\Gamma_{{j}^\star}
\left ( {\cal N}-{\bar {\cal N}}\right)_{\Gamma\Lambda}
\Bigl ( {\tilde {F}}^{\Lambda -}_{\mu\nu}  \nonumber\\
&&+ {1\over 8}
\nabla_{k}  f^{\Lambda}_{\ell} \bar \lambda^{kA}
\gamma_{\mu\nu} \, \lambda^{\ell B} \epsilon_{AB}
  \nonumber\\
&&    -
{1\over 4} \, \IC^{\alpha  \beta}\,{\bar\zeta}_{\alpha}\gamma _{\mu\nu}
\, \zeta _{\beta}\, L^{\Lambda}
\Bigr )
\nonumber \\
G^{{{i}^\star}+}_{\mu\nu} &=& {{\rm i} \over 2}\,g^{{{i}^\star}j} f^{\Gamma}_j
\left ({\cal N}-{\bar {\cal N}}\right)_{\Gamma\Lambda}
\Bigl ( {\tilde{F}}^{\Lambda +}_{\mu\nu} \nonumber\\
&& +{1\over 8}
\nabla _{{k}^\star} \bar f^{\Lambda}_{{\ell}^\star} \bar \lambda^{{k}^\star}_A
\gamma _{\mu\nu} \, \lambda^{{\ell}^\star}_B \epsilon^{AB}\nonumber\\
&&   - {1 \over 4}\,
\IC_{\alpha  \beta}\,{\bar\zeta}^{\alpha}\gamma _{\mu\nu} \,
\zeta ^{\beta}\, {\bar L}^{\Lambda}
\Bigr )\nonumber \\
D^{iAB} &=& {{\rm i} \over 2}g^{i{{j}^\star}}
C_{{{j}^\star}{{k}^\star}{{\ell}^\star}} \bar\lambda^{{k}^\star}_C
\lambda^{{\ell}^\star}_D
\epsilon^{AC} \epsilon^{BD}\,\nonumber\\
&& +\,W^{iAB}\nonumber\\
\end{eqnarray}
In the above equations we have denoted by ${\tilde{F}}$
the supercovariant field strength defined by:
\begin{eqnarray}
\tilde{F}^{\Lambda}_{\mu\nu} &=& {\cal F}^{\Lambda}_{\mu\nu}\,+
\,L^{\Lambda}\bar{\psi}^A_{\mu}\psi^B_{\nu}
\,\epsilon_{AB} \,\nonumber\\
&& +\bar L^{\Lambda}\bar{\psi}_{A \mu} \psi_{B \nu}\epsilon^{AB}\,
\nonumber\\
&& -{\rm i} \,f^{\Lambda}_i \,\bar {\lambda}^{iA} \gamma_{[\nu} \psi^B_{\mu]}\,\epsilon _{AB}\,
\nonumber\\
&&-{\rm i} \,{\bar f}^{\Lambda}_{{i}^\star} \,\bar\lambda^{{i}^\star}_A
\gamma_{[\nu} \psi _{B \mu]} \,\epsilon^{AB}\
\end{eqnarray}
\section{Final comments}
Let us make some observation about the structure of the Lagrangian
and of the transformation laws.
\par
i) We note that all the terms of the Lagrangian are given in terms of
purely geometric objects pertaining to either the special or the quaternionic
geometry. Furthermore the Lagrangian does not rely on the existence
of a prepotential function $F=F\left(X\right)$ and it is valid for
any choice of the quaternionic manifold.
\par
ii) The Lagrangian is not invariant under symplectic duality
transformations. However, in absence of gauging ($g=0$), if we
restrict the Lagrangian to configurations where the vectors are
on shell, it becomes symplectic invariant
\par
iii) We note that the field strengths ${\cal F}^{\Lambda\,-}_{\mu\nu}$ originally
 introduced in the Lagrangian are the free gauge field
strengths.The interacting field strengths which are supersymmetry
eigenstates are defined as the objects appearing in the
transformation laws of the gravitinos and gauginos
fields,respectively,namely the bosonic part of $T^-_{\mu\nu}$ and
$G^{-\,i}_{\mu\nu}$. In static configurations the integral of
these objects on a 2--sphere at infinity define the central charge
and the matter charges.

%%%%%%%%%%%%%%%%%%%%%%%%%%%%%%%%%%%%%%%%%%%%%%%%%%%%
% End of File action.tex %%%%%%%%%%%%%%%%%%%%%%%%%%%
%%%%%%%%%%%%%%%%%%%%%%%%%%%%%%%%%%%%%%%%%%%%%%%%%%%%

%%%%%%%%%%%%%%%%%%%%%%%%%%%%%%%%%%%%%%%%
\begin{table*}
\begin{center}
\caption{\sl Scalar Manifolds of Extended Supergravities}
\label{topotable}
\begin{tabular}{|c||c|c|c||c|c||c||c| }
\hline
\hline
~ & $\#$ scal. & $\#$ scal. & $\#$ scal. & $\#$ vect. &
 $\#$ vect. &~ & $~ $ \\
N & in & in & in & in  &
 in  &$\Gamma_{cont}$ & ${\cal M}_{scalar}$   \\
 ~ & scal.m. & vec. m. & grav. m. & vec. m. & grav. m. & ~ &~
\\
\hline
\hline
~    &~    &~   &~   &~  &~  & ~ & ~ \\
$1$  & 2 m &~   & ~  & n &~  &  ${\cal I}$  & ~   \\
~    &~    &~   &~   &~  &~  &  $\subset Sp(2n,\IR)$ & K\"ahler \\
~    &~    &~   &~   &~  &~  & ~ & ~ \\
\hline
~    &~    &~   &~   &~  &~  & ~ & ~ \\
$2$  & 4 m & 2 n& ~  & n & 1 &  ${\cal I}$ & Quaternionic $\otimes$
\\
~    &~    &~   &~   &~  &~  &  $\subset Sp(2n+2,\IR)$ & Special K\"ahler \\
~    &~    &~   &~   &~  &~  & ~ & ~ \\
\hline
~    &~    &~   &~   &~  &~  & ~ & ~ \\
$3$  & ~   & 6 n& ~  & n & 3 &  $SU(3,n)$ &~  \\
~    &~    &~   &~   &~  &~  & $\subset Sp(2n+6,\IR)$ & $\o{SU(3,n)}
{S(U(3)\times U(n))}$ \\
~    &~    & ~  &~   &~  &~  & ~ & ~ \\
\hline
~    &~    &~   &~   &~  &~  & ~ & ~ \\
$4$  & ~   & 6 n& 2  & n & 6 &  $SU(1,1)\otimes SO(6,n)$ &
$\o{SU(1,1)}{U(1)} \otimes $ \\
~    &~    &~   &~   &~  &~  & $\subset Sp(2n+12,\IR)$ &
$\o{SO(6,n)}{SO(6)\times SO(n)}$ \\
~    &~    &~   &~   &~  &~  & ~ & ~ \\
\hline
~    &~    &~   &~   &~  &~  & ~ & ~ \\
$5$  & ~   & ~  & 10 & ~ & 10 & $SU(1,5)$ & ~  \\
~    &~    &~   &~   &~  &~  & $\subset Sp(20,\IR)$ & $\o{SU(1,5)}
{S(U(1)\times U(5))}$ \\
~    &~    &~   &~   &~  &~  & ~ & ~ \\
\hline
~    &~    &~   &~   &~  &~  & ~ & ~ \\
$6$  & ~   & ~  & 30 & ~ & 16 & $SO^\star(12)$ & ~ \\
~    &~    &~   &~   &~  &~  & $\subset Sp(32,\IR)$ &
$\o{SO^\star(12)}{U(1)\times SU(6)}$ \\
~    &~    &~   &~   &~  &~  & ~ & ~ \\
\hline
~    &~    &~   &~   &~  &~  & ~ & ~ \\
$7,8$& ~   & ~  & 70 & ~ & 56 & $E_{7(-7)}$  & ~ \\
~    &~    &~   &~   &~  &~  & $\subset Sp(128,\IR)$ &
$\o{ E_{7(-7)} }{SU(8)}$ \\
~    &~    &~   &~   &~  &~  & ~ & ~ \\
\hline
 \hline
\end{tabular}
\end{center}
\end{table*}
\onecolumn{
\begin{center}
\begin{tabular}{ccccc}
\null & \null & {\large Table 2}: & \null & \null\\
\hline
\null & \null & \null & \null & \null\\
\null & \null & {\Large {\bf The N=2 Supergravity action}} & \null & \null \\
\null & \null & \null & \null & \null\\
\hline
\end{tabular}
\end{center}
\vskip 0.2cm
\begin{eqnarray*}
S=  \int \sqrt{-g}\,d^4 \,x \left ({\cal L}_{grav} \,+ \, {\cal L}_{kin} \,
+ \, {\cal L}_{Pauli} \,+\, {\cal L}_{gauging}\, + \,
{\cal L}_{4 ferm}
\right)
\end{eqnarray*}
\begin{eqnarray*}
{\cal L}_{grav} & = & - {1 \over 2} \, R \, + \,
\left (\bar {\Psi} ^A_{\mu} \gamma _{\sigma} \rho _{A \vert {\nu \lambda}} \,
- \, \bar {\Psi}_{A \mu} \gamma _{\sigma} \rho ^A_{\nu \lambda} \right )
\, {\epsilon^{\mu \nu \lambda \sigma} \over \sqrt{-g}}\nonumber %\label{gravaction} \\
\nonumber\\
{\cal L}_{kin} &= &  g_{i {{j}^\star}}\,
\nabla^{\mu} z^i \nabla _{\mu} \bar z^{{j}^\star}\,+\,
h_{uv} \, \nabla _{\mu}\, q^u \, \nabla^{\mu}\, q^v
\nonumber\\
&& - \,{{\rm i} \over 2}\,
g_{i{{j}^\star}}\, \left (\bar {\lambda}^{iA}\gamma ^{\mu} \nabla _{\mu} \lambda ^{{j}^\star}_A
\, + \,\bar {\lambda}^{{j}^\star}_A \gamma ^{\mu} \nabla _{\mu}\lambda ^{iA}\right ) \,
-\,{\rm i}\,
\, \left (\bar {\zeta}^{\alpha}\gamma ^{\mu} \nabla _{\mu}\zeta _{\alpha}
\, + \,\bar {\zeta}_{\alpha}\gamma ^{\mu} \nabla _{\mu}\zeta ^{\alpha} \right ) \nonumber \\
&& + \,{\rm i} \,\left(
\bar {\cal N}_{\Lambda \Sigma} {\cal F}^{- \Lambda}_{\mu \nu} {\cal F}^{- \Sigma \vert {\mu \nu}}
\, - \,
{\cal N}_{\Lambda \Sigma} {\cal F}^{+ \Lambda}_{\mu \nu} {\cal F}^{+ \Sigma \vert {\mu \nu}} \right )
 \nonumber %\label{kinaction}  \\
\nonumber\\
{\cal L}_{Pauli} \,&= & {\cal L}_{Pauli}^{inv} \,+
{\cal L}_{Pauli}^{non\,inv} \\
 \nonumber\\
{\cal L}_{Pauli}^{inv}\,&=&-\,g_{i{{j}^\star}}
\left( \nabla _{\mu} \bar z^{{j}^\star}\bar {\Psi}^{\mu}_A \lambda ^{i A}\, +\, h. c. \right)
- 2 \left( {\cal U}^{A \alpha}_{u}\,\nabla _{\mu}\, q^u \,
\bar {\Psi}_A^{\mu} \zeta _{\alpha} \,+\,h. c. \right) \nonumber\\
&& +\, g_{i{{j}^\star}} \left( \nabla _{\mu} \bar z^{{{j}^\star}} \,\bar {\lambda}^{iA}
\gamma^{\mu \nu} \Psi _{A \vert {\nu}}\,+\,h.c.\right)
+ \,2\, \left( {\cal U}^{\alpha A}_{\mu} \,\bar {\zeta}_{\alpha} \gamma^{\mu \nu}
\Psi _{A \vert {\nu}}\,+\,h.c. \right)
  \\
\nonumber\\
{\cal L}_{Pauli}^{non\,inv}\,&=&{\cal L}_{Pauli}^{-\,non\,inv} \,+
\,{\cal L}_{Pauli}^{+\,non\,inv}
\qquad \qquad  \left ( {\cal L}_{Pauli}^{+\,non\,inv} = \left ({\cal L}_{Pauli}^{-\,non\,inv} \right )^\star
\right )  \\
\nonumber\\
{\cal L}_{Pauli}^{-\,non\,inv} \,&= &  {\cal F}^{-\Lambda}_{\mu \nu} \left(
{\cal N} \, - \, \bar {\cal N} \right ) _{\Lambda \Sigma}
{\lbrack} -\,2\,{\rm i}\,L^{\Sigma } \bar {\Psi}^{A \vert \mu} \Psi ^{B \vert \nu} \epsilon _{AB}
\,-\,2 \,
{\bar f}^{\Sigma}_{{i}^\star}\,\bar {\lambda}^{{i}^\star}_A \gamma^{\nu} \Psi _B^{\mu}\,
\epsilon ^{AB}
\nonumber \\
&&- \, {{\rm i} \over 4}
\nabla _i f^{\Sigma}_j\,
\bar {\lambda}^{iA} \gamma^{\mu \nu} \lambda ^{jB} \,\epsilon _{AB} \, + \,
{{\rm i} \over 2}\,
L^{\Sigma} \bar {\zeta}_{\alpha} \gamma^{\mu \nu} \zeta _{\beta} \,C ^{\alpha \beta}
{\rbrack}  \\
\nonumber\\
{\cal L}_{gauging} &=& {\cal L}_{massmatrix} \, + \, {\cal L}_{potential} \\
\nonumber\\
{\cal L}_{massmatrix}
&=& \Biggl [ \,2\,g\, S_{AB} \bar {\Psi}^A_{\mu} \gamma^{\mu \nu}
\Psi ^B_{\nu} \, +
\,{\rm i}\,g\,g_{i{{j}^\star}}\,W^{iAB} \bar {\lambda}^{{j}^\star}_A
\gamma _{\mu} \Psi _B^{\mu}
\, + \, 2\, {\rm i} \,g\, N^A_{\alpha} \, \bar {\zeta}^{\alpha}
\gamma _{\mu} \Psi _A^{\mu} \nonumber\\
&& + \,
{\cal M}^{\alpha \vert \beta}\,
{\bar \zeta}_\alpha
\, \zeta_\beta \, + \, {\cal M}^{\alpha\vert }_{\phantom{\alpha}\vert iB}\,
{\bar \zeta}_\alpha \, \lambda^{iB} \, + \, {\cal M}_{iA\vert \ell B}
\, {\bar \lambda}^{iA} \lambda^{\ell
B}  \Biggr ] \, + \, \mbox{h.c.} \nonumber %\label{matricmass}\\
\nonumber\\
{\cal L}_{potential} &=& \,-{\rm V}\bigl ( z, {\bar
z}, q \bigr )=\nonumber\\
&& -\,g^2\,\Bigl[\left(g_{i{{j}^\star}} \, k^i_{\Lambda}\,k^{{j}^\star}_{\Sigma}\,+\,4\,h_{uv}
k^u_{\Lambda}\,k^v_{\Sigma} \right) \bar L^{\Lambda}\,L^{\Sigma}\nonumber\\
&& +\,g^{i{{j}^\star}}\,f^{\Lambda}_i\,f^{\Sigma}_{{j}^\star}\,
{\cal P}^x_{\Lambda}\,{\cal P}^x_{\Sigma}\,
-\,3\,\bar L^{\Lambda}\,L^{\Sigma}\,{\cal P}^x_{\Lambda}\,{\cal P}^x_{\Sigma}\Bigr]
\nonumber %\label{sugrapotential}\\
\nonumber\\
 {\cal L}_{4 ferm} &=& {\cal L}_{4 ferm}^{inv} \, + \,
 {\cal L}_{4 ferm}^{non \,inv}  \\
\nonumber\\
{\cal L}_{4 ferm}^{inv} & = & {{\rm i}\over 2} \left(
g_{i{{j}^\star}} \, \bar {\lambda}^{iA}\gamma _{\sigma}  \lambda ^{{{j}^\star}}_B \, - \,
2\,  \delta ^A_B \bar {\zeta}^{\alpha}\gamma _{\sigma} \zeta _{\alpha} \right )
\, \bar {\Psi}_{A \vert {\mu}} \gamma _{\lambda} \Psi ^B_{\nu}
{\epsilon^{\mu \nu \lambda \sigma} \over \sqrt{-g}}\nonumber \\
&& - \,{1 \over 6}\,
\left ( C_{ijk} \bar {\lambda}^{iA} \gamma^{\mu} \Psi ^B_{\mu} \,
\bar {\lambda}^{jC} \lambda ^{kD}\,
\epsilon _{AC} \epsilon _{BD} +h.c.\right)\nonumber \\
&&-2 \bar {\Psi}^A_{\mu} \Psi ^B_{\nu} \,\bar {\Psi}_A^{\mu} \Psi _B^{\nu}
+2g_{i{{j}^\star}}\,\bar {\lambda}^{iA} \gamma _{\mu} \Psi ^B_{\nu} \,
\bar {\lambda}^{{i}^\star}_A \gamma^{\mu} \Psi _B^{\nu}   \nonumber\\
&& + {1 \over 4}
\left (R_{i{{j}^\star}l{{k}^\star}}\,  + \,
g_{i{{k}^\star}} \, g_{l{{j}^\star}}\, -\, {3 \over 2}\,g_{i{{j}^\star}} \, g_{l{{k}^\star}}\right )
\bar {\lambda}^{iA}\lambda^{lB} \bar {\lambda}^{{j}^\star}_A\lambda^{{k}^\star}_B \,
\nonumber \\
&& +{1 \over 4} \,g_{i{{j}^\star}} \,
\bar {\zeta}^{\alpha} \gamma _{\mu} \zeta _{\alpha}\,
\bar {\lambda}^{iA} \gamma ^{\mu} \lambda^{{j}^\star}_A\,
+ \, {1 \over 2} \, {\cal R}^{\alpha}_{\beta ts}
\, {\cal U}^t_{A \gamma}\,{\cal U}^s_{B \delta} \epsilon ^{AB} \, C ^{\delta \eta}
\bar {\zeta}_{\alpha}\,\zeta _{\eta}\,\bar {\zeta}^{\beta}\,\zeta ^{\gamma} \nonumber \\
&& - \left[{{\rm i} \over 12} \nabla _m \, C_{jkl}
\bar {\lambda}^{jA}\lambda^{mB} \bar {\lambda}^{kC}\lambda^{lD}
\epsilon _{AC} \epsilon _{BD} +\,h.\,c.\right] \nonumber \\
&& + g_{i{{j}^\star}}\,
\bar {\Psi}^A_{\mu} \lambda ^{{j}^\star}_A  \,
\bar {\Psi}_B^{\mu} \lambda ^{i B}\,+\,
2  \bar {\Psi}^A_{\mu} \zeta ^{\alpha} \bar {\Psi}_A^{\mu} \zeta _{\alpha}\,\nonumber\\
&&+ \left ( \epsilon _{AB}\, \IC_{\alpha  \beta} \, \bar {\Psi}^A_{\mu} \zeta ^{\alpha} \,
  \bar {\Psi}^{B \vert \mu} \zeta ^{\beta} \,+ \,h. c.\right)
\nonumber %\label{4ferminv}  \\
\nonumber\\
{\cal L}_{4 ferm}^{non \,inv} & = & {\cal L}_{4 ferm}^{-\,non \,inv}
+{\cal L}_{4 ferm}^{+\,non \,inv} \qquad \qquad  \left (
{\cal L}_{4 ferm}^{+\,non \,inv} = \left ({\cal L}_{4 ferm}^{-\,non \,inv} \right )^\star
\right )  \\
\nonumber\\
{\cal L}_{4 ferm}^{-\,non \,inv} & = &
\left ({\cal N} - \bar {\cal N} \right )_{\Lambda \Sigma}\, \Bigl[
-\,{\rm i}\,L^{\Lambda}\, L^{\Sigma}\left( \bar {\Psi}^A_{\mu} \Psi ^B_{\nu}\right)^-
\left( \bar {\Psi}^C_{\mu} \Psi ^D_{\nu}\right)^- \epsilon _{AB} \, \epsilon _{CD} \nonumber\\
&&-\,4\,L^{\Lambda}{\bar f}^{\Sigma}_{{i}^\star}\left( \bar {\Psi}^A_{\mu} \Psi ^B_{\nu}\right)^-
\left(\bar {\lambda}^{{i}^\star}_A \gamma^{\nu} \Psi _B^{\mu}\right)^-\nonumber\\
&&+\,{\rm i}\,{\bar f}^{\Lambda}_{{i}^\star} {\bar f}^{\Sigma}_{{j}^\star}
\left(\bar {\lambda}^{{i}^\star}_A \gamma^{\nu} \Psi _B^{\mu}\right)^-
\left(\bar {\lambda}^{{j}^\star}_C \gamma _{\nu} \Psi _{D \vert \mu}\right)^-
\epsilon^{AB}\,\epsilon^{CD} \nonumber\\
&&+\,{1 \over 4}
L^{\Lambda}{\bar f}^{\Sigma}_{{\ell}^\star} \,g^{k \bar {\ell}}\,C_{ijk}
\left( \bar {\Psi}^A_{\mu} \Psi ^B_{\nu}\right)^-
\bar {\lambda}^{iC} \gamma^{\mu \nu} \lambda ^{jD}\,
\epsilon _{AB} \,\epsilon _{CD}
\nonumber \\
&& - \,{{\rm i} \over 2}
{\bar f}^{\Lambda}_{\bar m}{\bar f}^{\Sigma}_{{\ell}^\star} \,g^{k \bar {\ell}}\,C_{ijk}
\left(\bar {\lambda}^{\bar m}_A \gamma _{\nu} \Psi _{B \vert \mu}\right)^-
\bar {\lambda}^{iA} \gamma^{\mu \nu} \lambda ^{jB}\,
\nonumber \\
&& + {{\rm i} \over 2}
L^{\Lambda} L^{\Sigma}\left( \bar {\Psi}^A_{\mu} \Psi ^B_{\nu}\right)^-
\bar {\zeta}_{\alpha} \gamma^{\mu \nu} \zeta _{\beta} \,
\epsilon _{AB}\, \IC^{\alpha  \beta} \nonumber\\
 && + {1 \over 2} L^{\Lambda}{\bar f}^{\Sigma}_{{i}^\star}
\left(\bar {\lambda}^{{i}^\star}_A \gamma^{\nu} \Psi _B^{\mu}\right)^-
\bar {\zeta}_{\alpha} \gamma_{\mu \nu} \zeta _{\beta} \,
\epsilon^{AB} \IC^{\alpha  \beta}
\nonumber \\
&& + {{\rm i} \over 64}\,
C_{ijk}\,C_{lmn} g^{k{\bar r}} \, g^{n{\bar s}} \,
{\bar f}^{\Lambda}_{\bar r} \, {\bar f}^{ \Sigma}_{\bar s} \,
\bar {\lambda}^{iA} \, \gamma _{\mu \nu} \,\lambda^{jB} \,
\bar {\lambda}^{kC} \, \gamma^{\mu \nu}\,\lambda^{lD}
\, \epsilon _{AB} \epsilon _{CD}
\nonumber \\
&& + {{\rm i} \over 16}\,
L^{\Lambda} \nabla _i f^{\Sigma}_j
\bar {\zeta}_{\alpha} \gamma _{\mu \nu} \zeta _{\beta}\,
\bar {\lambda}^{iA} \gamma ^{\mu \nu} \lambda^{jB}\,
\,\epsilon _{AB} \,\IC^{\alpha  \beta}
\nonumber \\
&& - {{\rm i} \over 16}\,
L^{\Lambda} \,L^{\Sigma}
\bar {\zeta}_{\alpha} \gamma _{\mu \nu} \zeta _{\beta}\,
\bar {\zeta}_{\gamma} \gamma^{\mu \nu} \zeta _{\delta}\,
\IC^{\alpha  \beta}\,\IC^{\gamma \delta} \Bigr]
\nonumber %\label{4fermnoninv}
\end{eqnarray*}
}

\end{document}